\documentclass[12pt]{article}
\usepackage{amsfonts}
\usepackage{amssymb}
\usepackage{amsmath}
\usepackage{graphicx}

\renewcommand{\baselinestretch}{1.1}

\def\be{\begin{equation}}
\def\ee{\end{equation}}

\begin{document}

\begin{titlepage}

\begin{flushright}
hep-th/0412193\\
NSF-KITP-04-134\\
PUPT-2146\\
MIT-CTP-3577
\end{flushright}

\vspace{5mm}

\begin{center}
{\huge Cascading RG Flows from \\
\vspace{3mm}

New Sasaki-Einstein Manifolds}\\
\vspace{1mm}
\end{center}

\vspace{1mm}
\begin{center}
{\large C.~P.~Herzog$^\dagger$, Q.~J.~Ejaz$^*$, I.~R.~Klebanov$^\ddag$
}\\
\parbox{3in}{ \begin{center}
$\dag$ Kavli Institute for Theoretical Physics\\
University of California \\ Santa Barbara, CA  93106, USA\\
{\tt herzog@kitp.ucsb.edu} 
\end{center}}
\parbox{3in}{ \begin{center} $*$ Center for Theoretical Physics \\
Massachusetts Institute of Technology\\
Cambridge, MA 02139, USA\\
{\tt ejazqj@mit.edu}
\end{center}}\\
\vspace{2mm}
$\ddag$ Joseph Henry Laboratories\\ Princeton University, Princeton, NJ 08544, USA
\\
{\tt klebanov@princeton.edu}\\

\end{center}

\vspace{3mm}

\begin{center}
{\large Abstract}
\end{center}

\noindent
In important recent developments, 
new Sasaki-Einstein spaces $Y^{p,q}$ and conformal gauge
theories dual to $AdS_5\times Y^{p,q}$ have been constructed.
We consider a stack of $N$ D3-branes and $M$ wrapped D5-branes at
the apex of a cone over $Y^{p,q}$.
Replacing the D-branes by their fluxes, we construct asymptotic
solutions for all $p$ and $q$ 
in the form of warped products of the cone and $R^{3,1}$.
We show that they describe cascading RG flows where $N$ decreases
logarithmically with the scale. The warp factor, which we determine
explicitly, is a function of
the radius of the cone and one of the coordinates on $Y^{p,q}$.
We describe the RG cascades in the dual quiver gauge theories, and
find an exact agreement between the supergravity and the field theory
$\beta$-functions. We also discuss certain dibaryon operators and
their dual wrapped D3-branes in the conformal case $M=0$.

\vfil
\begin{flushleft}
December 2004
\end{flushleft}
\vfil
\end{titlepage}
\newpage

\renewcommand{\baselinestretch}{1.1}  

\section{Introduction}

An interesting generalization of the basic AdS/CFT correspondence
\cite{jthroat,US,EW} results from studying branes at conical
singularities \cite{ks,Kehag,KW,MP,Acharya}.  Consider a stack of $N$ D3-branes
placed at the apex of a Ricci-flat 6-d cone $Y_6$ whose base is a 5-d
Einstein manifold $X_5$. Comparing the metric with the D-brane
description leads one to conjecture that type IIB string theory on
$AdS_5\times X_5$ with $N$ units of 5-form flux,
is dual to the low-energy limit of the world volume
theory on the D3-branes at the singularity.

Well-known examples of $X_5$ are the orbifolds ${\bf S}^5/\Gamma$
where $\Gamma$ is a discrete subgroup of $SO(6)$ \cite{ks}. In these
cases $X_5$ has the local geometry of a 5-sphere.
Constructions of the dual gauge theories for Einstein manifolds $X_5$
which are not locally equivalent to ${\bf S}^5$ are also possible.
The simplest example is $X_5= T^{1,1}=
(SU(2)\times SU(2))/U(1)$ \cite{KW}.  The dual gauge theory is
the conformal limit of the world volume theory on a stack of $N$
D3-branes placed at the apex of
the conifold \cite{KW,MP}, which is a cone over $T^{1,1}$. This
${\cal N}=1$ superconformal gauge theory is
$SU(N)\times SU(N)$ with bifundamental fields
$A_i, B_j$, $i,j=1,2$, and a quartic superpotential.

Recently, a new infinite class of Sasaki-Einstein manifolds $Y^{p,q}$ of
topology $S^2\times S^3$ was discovered
\cite{Gauntlett2, Gauntlett}. 
Following progress in \cite{Martelli}, the ${\cal N}=1$ superconformal gauge theories
dual to $AdS_5\times Y^{p,q}$ were ingeniously constructed in
\cite{Benvenuti}. These quiver theories have gauge
groups $SU(N)^{2p}$, bifundamental matter, and marginal superpotentials
involving both cubic and quartic terms. These constructions generalize the 
SCFT on D3-branes placed at the apex of the complex cone 
over $dP_1$ \cite{Feng}, corresponding to $Y^{2,1}$ \cite{Martelli}. 
Impressive comparisons of the conformal anomaly coefficients
between the AdS and the CFT sides were carried out for $dP_1$
in \cite{Bertolini}, and in full generality in \cite{Benvenuti}.

In this paper we address a number of further issues concerning the
gauge/gravity duality involving the $Y^{p,q}$ spaces. We match 
the spectra of dibaryon operators in the gauge theory with that
of wrapped D3-branes in the string theory. Next, we consider
gauge theories that arise upon addition of $M$ wrapped D5-branes at
the apex of the cone. Our discussion generalizes that
given in \cite{chaotic, Hanwalls} for the $Y^{2,1}$ case.
We show that these gauge theories can undergo
duality cascades, and construct the dual warped 
supergravity solutions with $(2,1)$ flux.\footnote{
The duality cascade was first developed
for the conifold in \cite{KT, KS} and later generalized
in \cite{cascades, chaotic, Hanwalls}.
}

As a preliminary, in the next two sections 
we review the gauge theory duals for and the geometry
of these $Y^{p,q}$ spaces. 

\section{The Conformal Surface of $Y^{p,q}$ Gauge Theories}
\label{LeighStrassler}

In this section, we review the construction of the $Y^{p,q}$ gauge
theories and argue that they flow to an IR
conformal ``fixed surface'' of dimension two.  That this surface has
dimension  
two will be more or less clear from the gravity side where the two free
complex parameters are $C- i e^{-\phi}$ and
$\int_{S^2} (C_2 - i e^{-\phi} B_2)$.

\begin{figure}
\begin{center}
\includegraphics[width=2in]{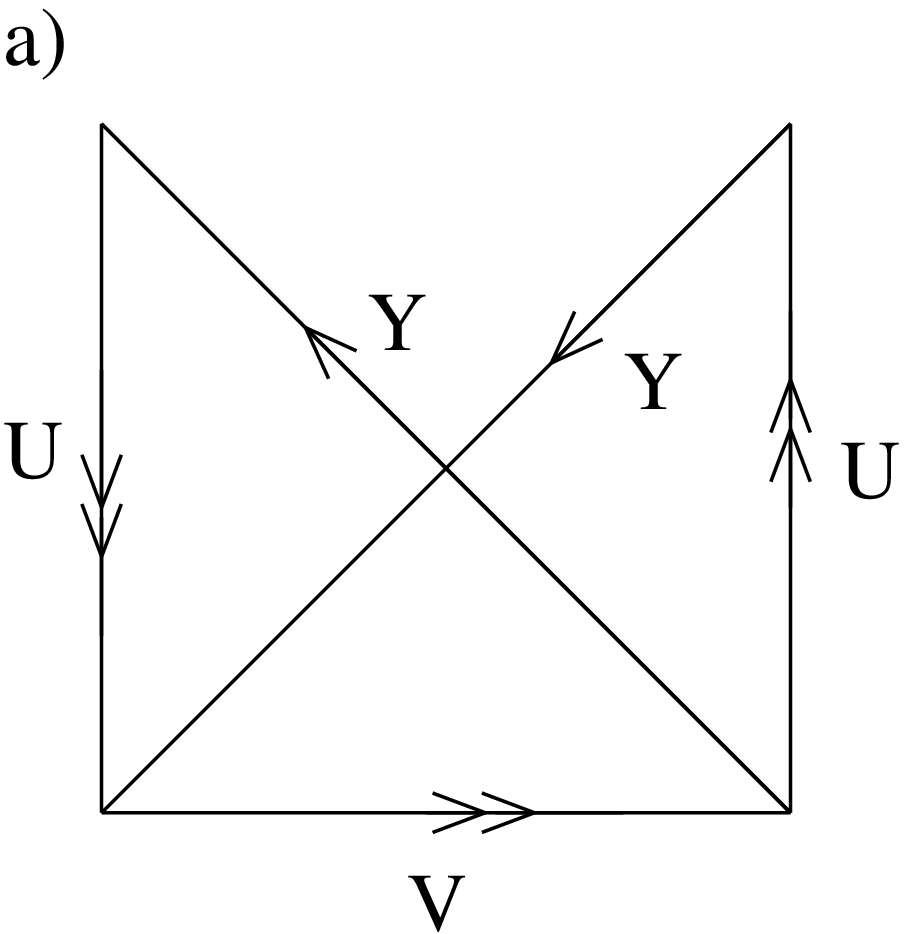}
\hfil
\includegraphics[width=2in]{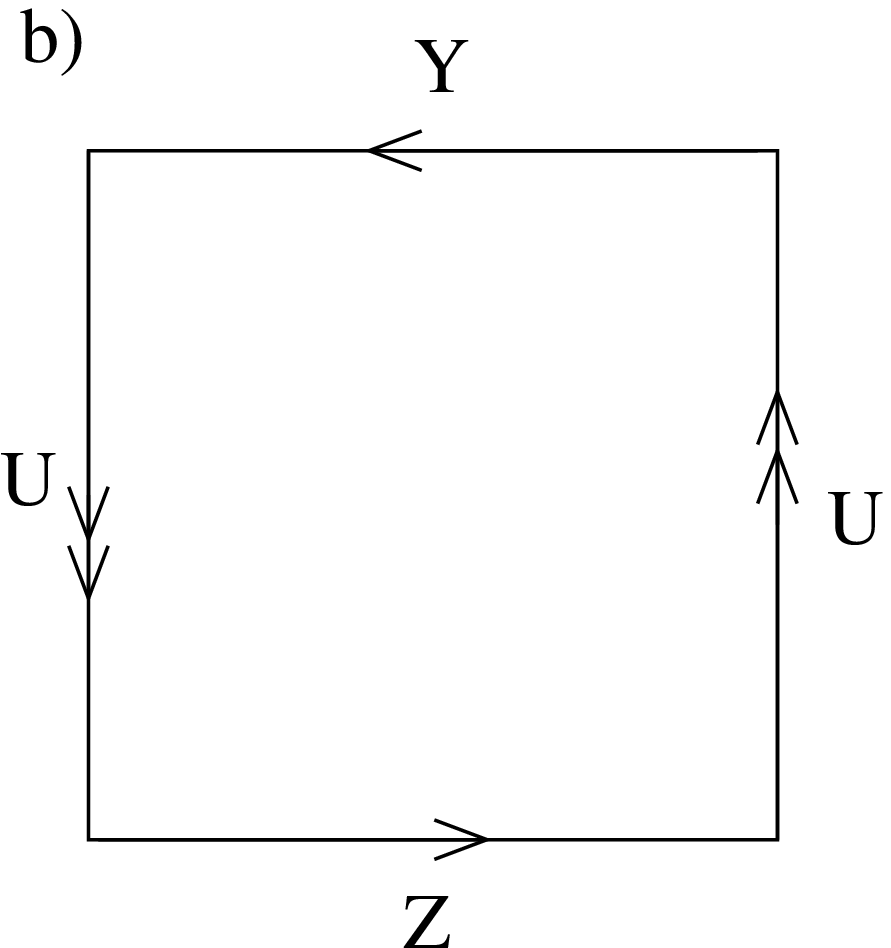} \\
\vfil
\includegraphics[width=4in]{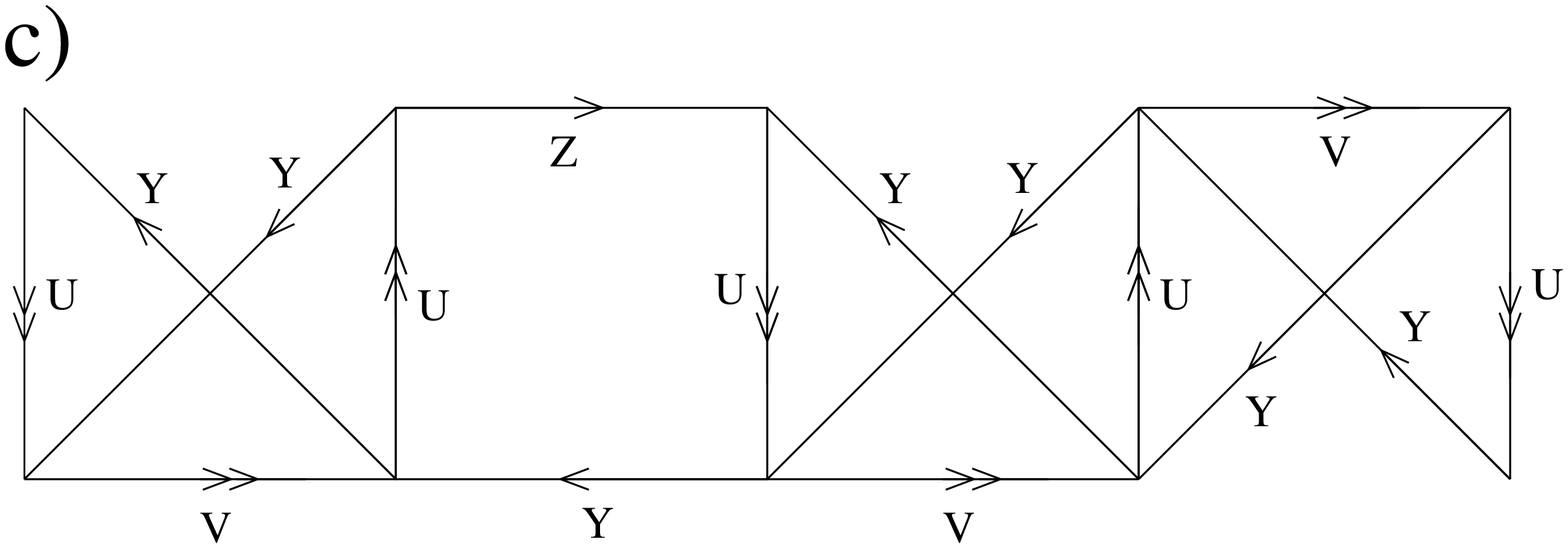}
\end{center}
\vfil
\caption{Shown are a) the unit cell $\sigma$; b) the unit cell $\tau$;
and c) the quiver for $Y^{4,3}$,
$\sigma \tilde \tau \sigma \tilde \sigma$.}
\label{unitcell}
\end{figure}

As derived in \cite{Benvenuti}, the quivers for these $Y^{p,q}$ gauge 
theories can be constructed from two basic units, $\sigma$ and $\tau$.
These units are shown in Figure \ref{unitcell}.
To construct a general quiver for $Y^{p,q}$, we define some basic
operations with $\sigma$ and $\tau$.  First, there are the inverted
unit cells, $\tilde \sigma$ and $\tilde \tau$, 
which are mirror images of $\sigma$ and
$\tau$ through a horizontal plane.  To glue the cells together,
we identify the double arrows corresponding to the $U^\alpha$
fields on two unit cells.  The arrows have to be pointing in 
the same direction for the identification to work.  So for
instance we may form the quiver $\sigma \tilde \tau = \tilde \tau \sigma$,
but $\sigma \tau$ is not allowed.  In this notation,
the first unit cell is to be glued not only to the cell on 
the right but also to the last cell in the chain.  A general 
quiver might
look like
\be
\sigma \tilde \sigma \sigma \tilde \tau \tau \tilde \sigma \ .
\ee
In general, a $Y^{p,q}$ quiver
consists of $p$ unit cells of which $q$ are of type $\sigma$.
The $Y^{p,p-1}$ gauge theories will have only one $\tau$ type 
unit cell, 
while the $Y^{p,1}$ theories will have
only one $\sigma$ type unit cell. 

Each node of the quiver corresponds to a gauge group while 
each arrow is a chiral field transforming in a bifundamental representation.
 For the $Y^{p,q}$ spaces, there are four types of bifundamentals
labeled $U^\alpha$, $V^\alpha$, $Y$, and $Z$ where $\alpha =1$ or 2.
To get a conformal theory, we take all the gauge groups to be $SU(N)$.
Later in this paper, when we add D5-branes, we will change
the ranks of some of the gauge groups and break the conformal symmetry.

The superpotential for this quiver theory is constructed by summing
over gauge invariant operators cubic and quartic in the fields
$U^\alpha$, $V^\alpha$, $Y$, and $Z$.  For each $\sigma$ unit cell
in the gauge theory, we add two cubic terms to the superpotential
of the form
\be
 \epsilon_{\alpha \beta} U^{\alpha}_L V^\beta Y  \; \; \mbox{and}
\; \;  \epsilon_{\alpha \beta} U^{\alpha}_R V^\beta Y \ .
\ee
Here, the indices $R$ and $L$ specify which group of $U^\alpha$ enter in the
superpotential, the $U^\alpha$ on the right side or the left side of $\sigma$.
The trace over the color indices has been suppressed.
 For each $\tau$ unit cell, we add the quartic term
\be
\epsilon_{\alpha\beta} Z U_R^\alpha Y U_L^\beta \ .
\ee

An analysis of the locus of conformal field theories 
begins with counting the fundamental
degrees of freedom which are in this case the $2p$ gauge couplings
and the $p+q$ superpotential couplings (assuming an unbroken
$SU(2)$ symmetry for the $U^\alpha$ and $V^\alpha$).  
We will assume all the gauge groups have
equal ranks.
There are in total
$3p+q$ fields and thus $3p+q$ anomalous dimensions which we
can tune to get a conformal theory.  We think of the $3p+q$
$\beta$-functions as functions of the $3p+q$ anomalous dimensions
which are in turn functions of the $3p+q$ coupling strengths,
$\beta_j(\gamma_i(g_k))$.

Let us check
that one set of solutions of $\beta_j=0$ involves setting
the anomalous dimensions of all the $Z$ fields equal, the anomalous
dimensions of all the $Y$ fields equal, and similarly for the $U^\alpha$
and $V^\alpha$.  Instead of working with the anomalous dimensions $\gamma$,
of
the fields, we find it convenient to work with the R charges, $R_Y$,
$R_Z$, $R_U$, and $R_V$.  (For superconformal gauge
theories, recall that $2(1+\gamma) = 3 R$.)

There are $p+q$ $\beta$-functions 
for the superpotential couplings.  
$p-q$ of the $\beta$ 
functions vanish when
$R_Z+R_Y+2R_U = 2$ and are associated with loops in the
$\tau$ unit cells, while the 
remaining $2q$ vanish when
$R_U+R_Y+R_V = 2$ and are associated to loops in the $\sigma$ unit cells.

There are $2p$ $\beta$-functions
for the gauge couplings.  
$2q$ of these couplings are associated 
with the
$\sigma$ unit cells, and the beta functions for these couplings vanish 
when $2=R_U+R_V+R_Y$ while the remaining 
$2p-2q$ belong to
the $\tau$ unit cells and vanish when $R_Z+R_Y+2R_U=2$.  Thus
the gauge coupling $\beta$-functions contain exactly the same information
as the superpotential $\beta$-functions.

It could be that there are more solutions to setting the 
$\beta_j=0$ which involve more generic values for the anomalous 
dimensions.  However, such solutions would require even more
degeneracy among the $3p+q$ $\beta$-functions, which is unlikely.  
Assuming we have found the most general solution of $\beta_j=0$
(which we have checked for $dP_1$ but should be checked in general),
we have found that only $3p+q-4+2$ of the $\beta_j=0$ are
linearly independent.  Thus there is seemingly 
a two dimensional plane
in the space of allowed anomalous dimensions which produce
conformal field theories.  Of course we know that $a$-maximization
\cite{IW}
will pick out the right anomalous dimensions.

However, there is a different way of looking at these $3p+q-2$
linearly independent $\beta$-functions.  They place
$3p+q-2$ constraints on the $3p+q$ couplings, leaving a 
space of conformal theories with two complex dimensions.
By construction, this space preserves the $SU(2)\times U(1)\times
U(1)$ global flavor symmetry of the $Y^{p,q}$.
If we allow a breaking of this symmetry, then there may exist
additional exactly marginal superpotential deformations (see
\cite{Leigh}).

\section{Review of the $Y^{p,q}$ geometry}

The $Y^{p,q}$ spaces are topologically $S^2 \times S^3$, and
the Sasaki-Einstein metric on them takes the form \cite{Gauntlett2, Gauntlett}
\begin{eqnarray}
d\Omega_{Y^{p,q}}^2 &=& \frac{1-y}{6} (d\theta^2 + \sin^2 \theta d\phi^2)
+ \frac{1}{w(y) v(y)} dy^2 + \frac{v(y)}{9} (d\psi -\cos \theta d\phi)^2
\nonumber \\
&& + w(y) [d\alpha + f(y)(d\psi - \cos\theta d\phi)]^2
\label{Ypqmetric}
\end{eqnarray}
where
\begin{eqnarray}
w(y) &=& \frac{2(b-y^2)}{1-y} \ , \\
v(y) &=& \frac{b-3y^2+2y^3}{b-y^2} \ , \\
f(y) &=& \frac{b-2y+y^2}{6(b-y^2)} \ .
\end{eqnarray}
 For the metric to be complete,
\be
b = \frac{1}{2} - \frac{p^2 - 3q^2}{4p^3}\sqrt{4p^2-3q^2} \ .
\ee
The coordinate $y$ is allowed to range between the two smaller roots
of the cubic $b-3y^2+2y^3$:
\begin{eqnarray}
y_1 &=& \frac{1}{4p}
\left( 2p - 3q - \sqrt{4p^2-3q^2} \right) \ , \\
y_2 &=& \frac{1}{4p}
\left( 2p + 3q - \sqrt{4p^2-3q^2} \right) \ .
\end{eqnarray}
The three roots of the cubic satisfy $y_1+y_2+y_3=3/2$, so the
biggest root, which we will need later in the paper, is
\be
y_3= 
\frac{1}{4p}
\left( 2p + 2 \sqrt{4p^2-3q^2} \right) \ .
\ee
The period of $\alpha$ is $2\pi \ell$ where
\be
\ell = - \frac{q}{4 p^2 y_1 y_2} \ .
\ee
The remaining coordinates are allowed the following ranges:
$0 \leq \theta < \pi$, $0 \leq \phi < 2\pi$, and $0 \leq \psi < 2\pi$.

The volume of $Y^{p,q}$ is given by
\be
\mbox{Vol}(Y^{p,q}) = \frac{q(2p+\sqrt{4p^2-3q^2}) \ell \pi^3}{3p^2} \ .
\ee

\section{Dibaryons and New 3-Cycles}

We will identify some new supersymmetric 3-cycles in the $Y^{p,q}$ geometry,
but first recall that Martelli and Sparks \cite{Martelli} 
identified two supersymmetric 3-cycles,
denoted $\Sigma_1$ and $\Sigma_2$ in their paper.  These three
cycles are obtained by setting $y=y_1$ or $y=y_2$ respectively.
At these values for $y$, the circle parametrized by $\psi$ shrinks
to zero size, and the three cycles can be thought
of as a $U(1)$ bundle parametrized by $\alpha$ over the 
round $S^2$ parametrized by $\theta$ and $\phi$.

Martelli and Sparks \cite{Martelli} 
computed the R-charges of the 
dibaryons corresponding 
to D3-branes wrapped on $\Sigma_1$ and $\Sigma_2$.  In general,
these R-charges are given by the formula \cite{GK}
\be
R(\Sigma_i) = \frac{\pi N}{3} \frac{\mbox{Vol}(\Sigma_i)}{\mbox{Vol}(Y^{p,q})} \ .
\ee
 From this general formula, it follows that
\begin{eqnarray} \label{RY}
R(\Sigma_1) &=& \frac{N}{3q^2} \left( 
-4p^2 + 2pq + 3q^2 + (2p-q) \sqrt{4p^2-3q^2} \right) \ , \\
\label{RZ}
R(\Sigma_2) &=& \frac{N}{3q^2} \left(
-4p^2 - 2pq + 3q^2 + (2p+q) \sqrt{4p^2-3q^2} \right) \ .
\end{eqnarray} 
These R-charges should correspond to operators ${\rm det} (Y)$
and $\rm{det} (Z)$ made out of the bifundamental fields
that are singlet under the global $SU(2)$ symmetry. Dividing
these dibaryon R-charges by $N$, we observe
a perfect match with the R-charges of the $Y$ and $Z$ singlet
fields determined from gauge theory
by Benvenuti, Franco, Hanany, Martelli, 
and Sparks \cite{Benvenuti}, $R_Y  = R(\Sigma_1)/N$ and
$R_Z = R(\Sigma_2)/N$.\footnote{The gauge theory
computation for $Y^{2,1}$  was performed earlier by
\cite{Bertolini}.}

Here we show which 3-cycles correspond to the dibaryons made 
out of the $SU(2)$ doublet
fields $U^\alpha$ and $V^\alpha$.\footnote{We would like to thank
S.~Benvenuti and J.~Sparks for discussions about these new 3-cycles.
A similar analysis will likely appear in a revision of \cite{Benvenuti}.} 
Such dibaryons carry spin $N/2$ under
the global $SU(2)$. On the string side, the wrapped D3-brane should
therefore have an $SU(2)$ collective coordinate (see \cite{GK}
for an analogous discussion in the case of $T^{1,1}$). 
The only possibility is that this $SU(2)$ is precisely
the $SU(2)$ of the round $S^2$ in the metric. Therefore, the 3-cycles
corresponding to these dibaryons should be localized at a point on the $S^2$. 

Now recall from the 
gauge theory analysis of \cite{Benvenuti}
that
\begin{eqnarray} \label{RU}
R_U &=& (2p(2p-\sqrt{4p^2-3q^2}))/3q^2 \ , \\
\label{RV}
R_V &=& (3q-2p+\sqrt{4p^2-3q^2})/3q \ .
\end{eqnarray}
Before proceeding, note that $R_V = R_U + R_Z$.
So if we determine which cycle $\Sigma_3$ corresponds to $U^\alpha$,
we can deduce that $V^\alpha$ is just a sum of $\Sigma_3$ and $\Sigma_2$.

As discussed above,
the three cycle $\Sigma_3$ 
should correspond to fixing a point on the $S^2$
and integrating over the fiber.  Setting $\phi = \theta = \mbox{const}$,
the induced metric on this three cycle becomes
\be
ds^2 = \frac{1}{wv} dy^2 + \frac{v}{9} d\psi^2 + w (d\alpha + f d\psi)^2 \ 
.
\label{sigma3met}
\ee

We can characterize this 3-cycle more precisely.
The metric on $\Sigma_3$ 
can be thought of as a principal $U(1)$ bundle over an $S^2$ where
the $S^2$ is parametrized by $y$ and $\psi$. 
A principal $U(1)$ bundle over $S^2$ is a Lens space $S^3 / {\mathbb Z}_k$
where $k$ is given by the first Chern class $c_1$ of the fibration.  
The $A = f d\psi$ is a connection one-form on the
$U(1)$ bundle.  
 Because $\alpha$
ranges from 0 to $2\pi \ell$, $dA = 2\pi c_1 / \ell$.  
Integrating $c_1$ over the
$S^2$ yields
\be
\int_{S^2} c_1 = \frac{f(y_2) - f(y_1)}{\ell} = -p \ .
\ee
In other words, our $\Sigma_3$
is the Lens space $S^3 / {\mathbb Z}_p$.
In \cite{Martelli}, $\Sigma_1$ and $\Sigma_2$
were identified as the Lens spaces
$S^3 / {\mathbb Z}_{p+q}$ and $S^3 / {\mathbb Z}_{p-q}$
respectively.

We find
\be
\mbox{Vol}(\Sigma_3) = \int \sqrt{g} \, dy \, d\alpha \, d\psi =
\frac{4\pi^2 \ell}{3}(y_2 - y_1) \ ,
\ee
where we have used the fact 
from (\ref{sigma3met}) that $\sqrt{g} = 1/3$.
Plugging into the formula for the R-charge, 
indeed $R(\Sigma_3) = N R_U$.

We now imagine that $V^\alpha$ corresponds to adding the cycles
$\Sigma_3$ and $\Sigma_2$ together.  Indeed, these two cycles
intersect along a circle at $y=y_2$.

We also check that $\Sigma_3$ is
a supersymmetric cycle, or in other words that
the form $\frac{1}{2} J \wedge J$, where $J$ is the Kaehler
form on the cone over $Y^{p,q}$, restricts to the induced
volume form on the cone over $\Sigma_3$.  
More formally, we are checking that $\Sigma_3$ is calibrated by
$\frac{1}{2} J \wedge J$.

 From Martelli and Sparks (2.24) \cite{Martelli}, we find that
\begin{eqnarray*}
J &=& r^2 \frac{1-y}{6} \sin\theta \, d\theta \wedge d\phi \\
&& + \frac{1}{3} r \, dr \wedge (d\psi - \cos \theta \, d\phi)
-d(yr^2) \wedge \left(
d\alpha + \frac{1}{6} (d\psi - \cos \theta \, d\phi) \right) \ , 
\end{eqnarray*}
and hence
\be
\left. J \right|_{\Sigma_3}
= \frac{1}{3} r dr \wedge d\psi + d(yr^2) \wedge 
\left( d\alpha + \frac{1}{6} 
d\psi \right) \ .
\ee
Thus we find that
\be 
\left. 
\frac{1}{2} J \wedge J 
\right|_{\Sigma_3}
= \frac{r^3}{3} dr \wedge d\psi \wedge dy \wedge 
d\alpha 
\ee
as expected.

\section{Warped Solutions with (2,1) Flux}

The first step in constructing supersymmetric warped
solutions for these $Y^{p,q}$ spaces is constructing a harmonic
$(2,1)$ form $\Omega_{2,1}$.
We begin by rewriting the metric so that locally we have
a $U(1)$ fiber over a Kaehler-Einstein manifold.  From
(2.17) of \cite{Martelli}, we have
\be
d\Omega_{Y^{p,q}}^2 = (e^\theta)^2 + (e^\phi)^2 + (e^y)^2 + (e^\beta)^2
+ (e^\psi)^2
\ee
where we have defined the one forms
\be
e^\theta = \sqrt{\frac{1-y}{6}} d\theta \; , \; \; \;
e^\phi = \sqrt{\frac{1-y}{6}} \sin\theta d\phi \ , 
\ee
\be
e^y = \frac{1}{\sqrt{wv}}dy \; , \; \; \; 
e^\beta = \frac{\sqrt{wv}}{6} (d\beta + \cos\theta d\phi) \ ,
\ee
\be
e^\psi = \frac{1}{3} (d\psi - \cos\theta d\phi + y(d\beta + \cos\theta 
d\phi)) \ .
\ee
In terms of the original coordinates $\beta = -6 \alpha - \psi$.
Here, the $\psi$ is a coordinate on the local $U(1)$ fiber.

There is then a local Kaehler form, denoted 
$J_4$ by \cite{Martelli}, on the Kaehler-Einstein base:
\be
J_4 = e^\theta \wedge e^\phi + e^y \wedge e^\beta  .
\ee 
Based on \cite{chaotic}, we expect to
be able to construct $\Omega_{2,1}$ from a $(1,1)$ form $\omega$
using this local Kaehler-Einstein metric such that
$*_4 \omega = -\omega$, $d\omega =0$, and $\omega \wedge J_4 = 0$.
We guess that
\be
\omega = F(y) (e^\theta \wedge e^\phi - e^y \wedge e^\beta) \ .
\ee
The form $\omega$ is clearly anti-selfdual and orthogonal
to $J_4$. Using a complex basis of one-forms
constructed in (2.27) of
\cite{Martelli}, it is not hard to check that
$\omega$ is indeed a $(1,1)$ form. The condition $d\omega = 0$ then implies that
\be
F(y) = \frac{1}{(1-y)^2} \ .
\ee
Further, we construct a $(2,1)$ form from the wedge product of
a $(1,0)$ form and $\omega$:
\be
\Omega_{2,1} = K \left( \frac{dr}{r} + i e^\psi \right) \wedge \omega \ .
\ee
We have introduced a normalization constant $K$ for later convenience.
We have checked that $d\Omega_{2,1} = 0$ and $*_6 \Omega_{2,1} =  i 
\Omega_{2,1}$. 

Next, we analyze
\be
\int_{\Sigma_i} \Omega_{2,1}
\ee
for the three three-cycles $i=1$, 2, 3.  We find that
\begin{eqnarray}
\int_{\Sigma_1} \Omega_{2,1} &=& -K \frac{8i \pi^2 \ell}{3} 
\frac{y_1}{1-y_1} 
\ , \\
\int_{\Sigma_2} \Omega_{2,1} &=& -K \frac{8i \pi^2 \ell}{3} 
\frac{y_2}{1-y_2}
\ , \\
\int_{\Sigma_3} \Omega_{2,1} &=& -K \frac{4i \pi^2 \ell}{3}
\left( \frac{1}{1-y_2} - \frac{1}{1-y_1} \right) \ .
\end{eqnarray}

The ratios between these integrals look superficially to
be irrational.  However, the ratios must be rational, and
we find that if we set
\be
K = \frac{9}{8\pi^2} (p^2 - q^2)
\ee
then
\begin{eqnarray}
\int_{\Sigma_1} \Omega_{2,1} &=& -i(-p+q) \ , \\
\int_{\Sigma_2} \Omega_{2,1} &=& -i(p+q) \ , \\
\int_{\Sigma_3} \Omega_{2,1} &=& -ip \ .
\end{eqnarray}

Now, to construct a supergravity solution, we take the real
RR $F_3$ and NSNS $H_3$ forms to be
\be
i K' \Omega_{2,1} = F_3 + \frac{i}{g_s} H_3 \ ,
\ee 
\be
F_3 = - KK' e^\psi \wedge \omega \; ; \; \; \;
H_3 = g_s KK' \frac{dr}{r} \wedge \omega \ ,
\ee
where we have introduced another normalization constant $K'$.
In particular, $F_3$ should be quantized such that
\be
\int_{\Sigma_1} F_3 = 4 \pi^2 \alpha' M (p-q)
\ee
where $M$ is the number of D5-branes.  Thus we find that
$K' = 4 \pi^2 \alpha' M$.  (See \cite{review} for our 
normalization conventions.)

\subsection{Derivation of Five-Form Flux}

 For the metric and $F_5$ we take the usual ansatz with the warp factor
$h$,
\be ds^2= h^{-1/2} dx_4^2 + h^{1/2} (dr^2 + r^2 d\Omega_{Y^{p,q}}^2)
\ ,
\ee
\be \label{warpfive}
g_s F_5 = d(h^{-1})\wedge d^4 x + * [ d(h^{-1})\wedge d^4 x ] \ .
\ee
Due to the appearance of the $y$-dependent factor $F(y)$ in
the $(2,1)$ flux, it is inconsistent to assume that
$h$ is a function of $r$ only. 
Instead, similar
to the gravity duals of fractional branes on the ${\mathbb Z}_{2}$ orbifold
\cite{Bert},
$h$ is a function of two variables, $r$ and $y$.
 For $q\ll p$ the $y$-dependence
can be ignored, and the warp factor approaches that
found for the warped conifold in \cite{KT}.
On the other hand, for $p-q \ll p$ we find that $h$ gets sharply peaked
near $y=1$, and the solutions approach
the gravity duals of fractional branes in 
orbifold theories
\cite{KN,Joe,Bert}.
Thus, the warped solutions we find with the $Y^{p,q}$ serve
as interesting interpolations between the
conifold and the orbifold cases.


More explicitly, the first term in (\ref{warpfive}) is
\be
-h^{-2}\left ({\partial h \over\partial r} dr + \sqrt{wv} 
{\partial h \over\partial y} e^y \right ) \wedge d^4 x
\ .\ee
Working out its Hodge dual, and substituting into the equation
\be
dF_5= H_3\wedge F_3
\ ,\ee
we find the second order PDE
\be \label{genpde}
- (1-y) {\partial\over \partial r} \left (r^5 {\partial h \over\partial r}
\right ) -r^3 {\partial\over \partial y}
\left ((1-y) w v  {\partial h \over\partial y}\right )
= {C \over r (1-y)^3 }
\ ,
\ee
where $C \equiv 2 (g_s KK')^2$.
Note that, after dividing the PDE by $r^5 (1-y)$, we obtain the standard equation
\be
-\nabla_{pq}^2 h ={1\over 6} |H_3|^2
\ee
where $\nabla_{pq}^2$ is the Laplacian on the cone over $Y^{p,q}$.  

The supergravity lore predicts that supersymmetric solutions should obey
first order systems of differential equations.  Our supergravity solution,
based on $\Omega_{2,1}$ is expected to be supersymmetric if it has no
curvature singularities \cite{Dasgupta}.  Naively, this first order system
could be easier to solve than the second order PDE (\ref{genpde}).
Such a first order system for $F_5$ can be generated starting from the ansatz
\be
F_5 = B_2 \wedge F_3 + dC_4 
\ee
where
\be
g_s C_4 = h(r,y)^{-1} d^4x +  \frac{f(r,y)}{(1-y)\sqrt{wv}} e^\psi \wedge e^\theta \wedge
e^\phi \wedge e^\beta  \ ,
\ee
and we have used (\ref{NSNSB2}).
Enforcing the selfdual constraint $F_5 = *F_5$, one finds
\begin{eqnarray*}
\frac{\partial h}{\partial r} r^5 &=& \frac{\partial f}{\partial y} \frac{1}{1-y} +
\frac{C}{(1-y)^4} \ln r \ , \\
\frac{\partial h}{\partial y} r^3 (1-y) wv &=& -\frac{\partial f}{\partial r} \ ,
\end{eqnarray*}
which is indeed a first order system (a similar type of
system appears in a somewhat different context in \cite{Halmagyi}). 
Unfortunately, as it involves one more
function than our PDE (\ref{genpde}), it seems no easier to solve; in
fact this system is equivalent to (\ref{genpde}) as a constraint on $h(r,y)$.

\subsection{Solving for the Warp Factor}

First, we discuss the boundary conditions at $y=y_1$ and $y=y_2$.
At these points the radius of the circular coordinate $\psi$
smoothly shrinks to zero. Defining the coordinate
$\rho\sim \sqrt{y-y_1}$ near the boundary, we find that the metric in these
two dimensions (with other coordinates fixed) is locally
\be
ds_2^2= d\rho^2 + \rho^2 d\psi^2 \ .
\ee
The behavior of $\psi$-independent modes in these
radial coordinates is well-known. The boundary condition is
${dh\over d\rho} =0$, so that 
\be h= h_0 + h_2 \rho^2 + \ldots= h_0 +\tilde h_2 (y-y_1) + \ldots
\ .
\ee
In terms of the $y$-coordinate, we have the boundary conditions
that ${\partial h\over \partial y}$ is finite at the boundaries, while
$h$ is positive there.

Let us substitute into (\ref{genpde})
\be
h=r^{-4} f(t, y)\ ,\qquad t=\ln (r/r_0)
\ .
\ee
The PDE for $f(t,y)$ assumes the simpler form
\be
(1-y)\left (-{\partial^2 f\over \partial t^2} + 
4 {\partial f\over \partial t} \right )
-{\partial\over \partial y}\left (2(b-3 y^2 + 2 y^3) 
{\partial f\over \partial y}\right )= {C\over (1-y)^3}
\ .
\ee
Now, it is clear that there are solutions of the form
\be
f(t, y)= At + s(y)
\ ,
\ee
where $A$ is a constant,
and the ODE for $s(y)$ is
\be\label{genode}
-{d\over d y}\left (2(b-3 y^2 + 2 y^3) 
{ds\over dy}\right )= {C\over (1-y)^3} - 4 A (1-y)
\ .
\ee
The boundary conditions are that $s'$ is finite at both end-points.
Therefore, integrating the LHS from $y_1$ to $y_2$ we must find zero.
This imposes a constraint on $A$ that
\be
\int_{y_1}^{y_2} dy \left [ {C\over (1-y)^3} - 4 A (1-y)\right ]=0\ ,
\ee
whose solution is
\be
A = \frac{C}{4 (1-y_1)^2 (1-y_2)^2}
\ .\ee
Now we can integrate (\ref{genode}) twice to find
\be
s(y) = -\frac{C}{4(b-1)} \left[
\frac{1}{1-y} +
\frac{(1+2y_1)(1+2y_2) \ln(y_3-y)}{2(b-1)}\right ] + {\rm const}
\ .
\ee
This function has singularities at $y=y_3$ and $y=1$, but they are safely outside 
the region $y_1<y< y_2$ for all admissible $p$ and $q$.
To summarize, the warp factor we find is
\be
h(r,y) = {A\ln(r/r_0) + s(y)\over r^4 }
\ .
\ee
Just like the solution found in \cite{KT}, this solution has a naked
singularity for small enough $r$. It should be interpreted as the asymptotic
form of the solution. In the conifold case, the complete solution
\cite{KS} involves the deformation of the conifold that is important
in the IR, but in the UV the solution indeed approaches
the asymptotic form found earlier in \cite{KT}.
Finding the complete solutions for cones over $Y^{p,q}$, non-singular
in the IR, remains an important problem.

There are two 
interesting special limits of our solutions.
 For $q\ll p$, 
\be
y_1 =- {3q\over 4p} + O(q^2/p^2)
\ ,\qquad
y_2 = {3q\over 4p} + O(q^2/p^2) \ .
\ee
In this limit the range of $y$ becomes narrow, and both end-points approach zero.
Since ${\partial h\over \partial y}$ is finite, the variation 
of $h$ in the $y$-direction can be ignored, and
we have $h\sim \ln (r/r_0)/r^4$, as in \cite{KT}.
This is not surprising, since for $q\ll p$ the spaces $Y^{p,q}$
may be approximated by a ${\mathbb Z}_p$ orbifold of $T^{1,1}$.

The other special case is $q=p-l$, with $l\ll p$. Now
\be
b= 1- {27 l^2\over 4 p^2} + O(l^3/p^3)\ ,
\ee
and
\be
y_1 =- {1\over 2} + {3 l^2\over 2 p^2} + O(l^3/p^3)
\ ,\qquad
y_2 = 1- {3l\over 2 p}  + O(l^2/p^2)\ ,
\qquad
y_3= 1+ {3l\over 2 p}  + O(l^2/p^2)\ .
\ee
Note that $y_2$ approaches $1$ from below, while $y_3$ from above,
as ${l\over p}\to 0$.
In this limit,
we find that $h$ depends on $y$ strongly and gets sharply peaked
near $y=1$. While ${\partial h\over \partial y}$
is finite at $y_2$ for any finite $l$ and $p$,
it diverges in the limit $l/p\to 0$.
The limiting form of the warp factor is
\be
h(r, y)\rightarrow
6 {(\alpha' g_s M)^2 p^4\over r^4}\left [
{4\over 3} \ln (r/r_0) + {1\over 1-y} - {2\over 3} \ln (1-y)\right ]
\label{limith}
\ .
\ee
To facilitate comparison with the solution 
found for the $S^5/{\mathbb Z}_{2}$ orbifold case in \cite{Bert},
it is convenient to introduce a new coordinate $\rho$
\be
\frac{2}{3} (1-y) 
= 1 - \frac{\rho^2}{r^2} \ .
\ee
 For $q=p$ the variable 
$\rho$ ranges from from $0$ to $r$.  We also introduce
an auxiliary radial variable $r' = \sqrt{r^2 - \rho^2}$.

The geometry
of $Y^{p,p}$ is that of 
the ${\mathbb Z}_{p}$ orbifold of $S^5/{\mathbb Z}_{2}$.
%
In \cite{Gauntlett}, the space $Y^{1,1}$ was identified with 
the ${\mathcal N}=2$ preserving
$S^5/{\mathbb Z_2}$ orbifold.  In the limit $q \to p$,
the metric (\ref{Ypqmetric}) is independent of both $p$ and $q$.
Only the period of the $U(1)$ fiber coordinate 
$\alpha$, which becomes $\pi/p$
in this limit, depends on $p$.  
In the limit $p=q$, we can rewrite the metric on the cone over
(\ref{Ypqmetric}), $dr^2 + r^2 d\Omega_{Y^{p,p}}^2$, 
in the form
\be
ds^2= dr'^2 +  {1\over 4} r'^2
\left [ d\theta^2 + \sin^2 \theta \, d\phi^2 +
\left (-d \psi - 2 d\alpha + 
\cos \theta \, d\phi \right )^2\right ]
+ d\rho^2 + 4 \rho^2 (d\alpha)^2
\ .
\ee

 From this form of the metric, one can see that the cone
over $Y^{p,p}$ factors into a cone over an 
orbifolded $S^3$ and a cone over an 
orbifolded $S^1$. 
The cone over $S^3$ is locally ${\mathbb C}^2$
parametrized by $\theta$, $\phi$, $\psi + 2\alpha$,
and the auxiliary radial coordinate $r'$. 
In this Euler angle parametrization,
$-\psi - 2\alpha$ gives the overall phase of
$(z_1, z_2) \in {\mathbb C}^2$.
The cone over $S^1$ is parametrized by the angle
$\alpha$ and the radial coordinate $\rho$.

The orbifold action sends $\alpha \to \alpha - \pi/p$, acting
as ${\mathbb Z}_p$ on this cone over $S^1$.
If the range of $\alpha$ ran from zero to $\pi$ instead of from
zero to $\pi/p$, then the cone over $S^1$ would be smooth.
This same action shifts the phase
$\psi + 2\alpha \to \psi + 2\alpha - 2\pi/p$.
 For the $S^3$ to be unorbifolded, the Euler
angle $-\psi - 2\alpha$ should run from
zero to $4\pi$.  We conclude the orbifold
acts on this cone over $S^3$ as
${\mathbb Z}_{2p}$.

Putting the two cones together we find the
${\mathbb Z}_{2p}$ orbifold
of ${\mathbb C}^3$ described in
\cite{Benvenuti}.  More precisely, 
the cone over $Y^{p,p}$ is the orbifold generated by
$\zeta: (z_1, z_2, z_3) \to (\omega^{a_1} z_1,
\omega^{a_2} z_2, \omega^{a_3} z_3)$,
where $\omega$ is a $2p$'th root of unity and,
keeping track of the signs of the angles,
$\vec a = (1,1,-2)$.  We see that
$\zeta^p$ generates a ${\mathbb Z}_2$
subgroup of ${\mathbb Z}_{2p}$.  Moreover,
$\zeta^p$ acts as the identity on 
$z_3$, fixing a circle in $Y^{p,p}$.

 From this discussion,
$y=1$ 
(or equivalently $\rho = r$) 
is the 
location of the circle fixed by
$\zeta^p$.
In terms of the coordinate $\rho$, 
our warp factor (\ref{limith}) becomes 
\be
h(r, \rho)\rightarrow
4 {(\alpha' g_s M)^2 p^4\over r^4}\left [
\ln\left (r^4\over r^2- \rho^2\right ) + {r^2\over r^2-\rho^2} 
+ {\rm const}\right ]
\label{limithredef}
\ .
\ee
This matches
the warp factor (44) of \cite{Bert} exactly.


\section{Matching the $\beta$ Function}

In this section we match the supergravity and gauge theory calculations
of the beta function. On the supergravity side,
we can calculate the running of the gauge coupling constant $g$ on the 
stack of D5-branes from the integral of $B_2$ (recall $dB_2 = H_3$) 
over a two-cycle.
In particular
\be
\frac{8 \pi^2}{g^2} = \frac{1}{2\pi \alpha' g_s} \int_{\mathcal C} B_2 \ .
\ee
Now
\be
B_2 = (\ln r) (4 \pi^2 \alpha' g_s M) K \omega \ .
\label{NSNSB2}
\ee
It is unclear how to describe
the two-cycle ${\mathcal C}$ in terms of the metric coordinates.  However,
based on \cite{chaotic}, we expect that
the harmonic form Poincare 
dual to ${\mathcal C}$ is $K \, e^\psi \wedge \omega$.
Thus, we take
\be
K \int_C \omega = K^2 \int_{Y^{p,q}} 
e^\psi \wedge \omega \wedge \omega 
\ .
\ee
One quickly finds
\be
K \int_C \omega = \frac{p^2}{2\pi} 
\left( p + \sqrt{4p^2-3q^2} \right)
\ee
and hence that
\be
\frac{8\pi^2}{g^2} = (\ln r) M p^2 \left( p + \sqrt{4p^2-3q^2} \right) \ . 
\label{betageom}
\ee

On the gauge theory side, we expect that
\be
\beta_{D5} =  \sum s^i \beta_i  
\ee
where the vector $s^i$ describes how adding a D5-brane changes
the ranks of the gauge groups.  
In \cite{chaotic}, it was demonstrated that a cubic anomaly involving
the R and $U(1)_B$ charges is
related in a precise way to this particular weighted sum
of $\beta$ functions:
\be
\mbox{tr} R U(1)_B^2 = -\frac{2}{3M} \sum s^i \beta^i \ .
\ee
In the derivation of this formula, it was assumed that the
anomalous dimensions of the chiral fields are determined
by the R-charges of the conformal theory.  In principle,
there could be coupling constant corrections to these anomalous
dimensions if we start at a point away from the conformal
surface described in Section \ref{LeighStrassler} and then add
D5-branes.  Even if we start on the conformal surface, the
addition of D5-branes could conceivably introduce $M/N$
corrections to these anomalous dimensions.  The fact
that the geometric and gauge theory calculations will agree indicates
that these corrections should begin at order $(M/N)^2$, as discussed
in \cite{chaotic}.

Using R-charges for the chiral fields (\ref{RY}), (\ref{RZ}), (\ref{RU}), and (\ref{RV}), 
which were first derived for $Y^{2,1}$ in \cite{Bertolini} and later
for all $Y^{p,q}$ in  \cite{Benvenuti} using 
$a$-maximization, we can compute
\begin{eqnarray*}
\mbox{tr} R U(1)_B^2 &=& (p-1) (R_Z-1)(p+q)^2 + 2p(R_U-1)(-p)^2 \\
&& + \, 2q(R_V-1)q^2 + (p+q)(R_Y-1)(p-q)^2 \\
&=& -\frac{2}{3} p^2 \left( p + \sqrt{4p^2 - 3q^2} \right) \ ,
\end{eqnarray*}
which agrees with the intersection calculation above.

This calculation seems like a bit of magic.  As part of a more general
discussion of Seiberg duality \cite{Seibergduality}
cascades, we will repeat this calculation
using 
brute force for two classes $Y^{p,p-1}$ and $Y^{p,1}$ of spaces.

\section{Cascades in the Dual Gauge Theories}

The simplest example of a Seiberg duality cascade occurs in the
$SU(N+M)\times SU(N+2M)$ gauge theory with bifundamental fields
$A_i, B_j$, $i,j=1,2$, and a quartic superpotential \cite{KT,KS}. 
(The theory with $M=0$ is conformal -- the addition of the $M$
D5-branes breaks the conformal symmetry.)  In this case
the gauge coupling of $SU(N+2M)$ blows up after a finite amount of RG
flow. To continue the flow beyond this point, one applies the
duality transformation to this gauge group \cite{KS}. After this transformation,
and an interchange of the two gauge groups,
the new gauge theory is $SU(\tilde N + M)\times SU(\tilde N+2 M)$
with the same matter and superpotential, with $\tilde N=N-M$.
This self-similar structure of the gauge theory under Seiberg
duality is the crucial fact that allows the cascade to happen.
If $N= k M$, where $k$ is an integer, then the cascade stops after $k$
steps, and we find $SU(M)\times SU(2M)$ gauge theory. This IR
gauge theory exhibits a multitude of interesting effects visible
in the dual supergravity background, such as confinement,
and chiral symmetry breaking \cite{KS}. 
Particularly interesting
is the apearance of an entire ``baryonic branch'' of the moduli space
in the gauge theory \cite{KS,Aharony}, whose existence in the dual supergravity
was recently confirmed in \cite{GHK,Butti}.
The presence of the baryonic operators in the IR
gauge theory is related to the fact that
for the $SU(2M)$ gauge group, the number of flavors equals the number of colors.

The self-similar structure of the gauge theory under the duality, which allows
the cascade to occur, can be found
in more complicated quiver diagrams as well. In \cite{chaotic,Hanwalls}
the cascade in the gauge theory dual to $AdS_5\times Y^{2,1}$ was analyzed.
The relevant gauge theory is
$SU(N+M)\times SU(N+3M)\times SU(N+2M)\times SU(N+4M)$. If the initial
conditions are such that
the biggest gauge group flows to infinite coupling first, then
after applying a duality transformation to this group and permuting
factor groups, we find exactly the same theory, with $N\to N-M$.
 For a generic choice of initial conditions, 
the biggest gauge group will flow to infinite coupling again, 
and the cascade repeats
until $N$ reaches zero far in the infrared.

In fact, this structure of the cascade is possible for all gauge theories
dual to $AdS_5\times Y^{p,p-1}$ and $AdS_5 \times Y^{p,1}$. 

\subsection{Cascades for $Y^{p,p-1}$}

\begin{figure}
\begin{center}
\includegraphics[width=4in]{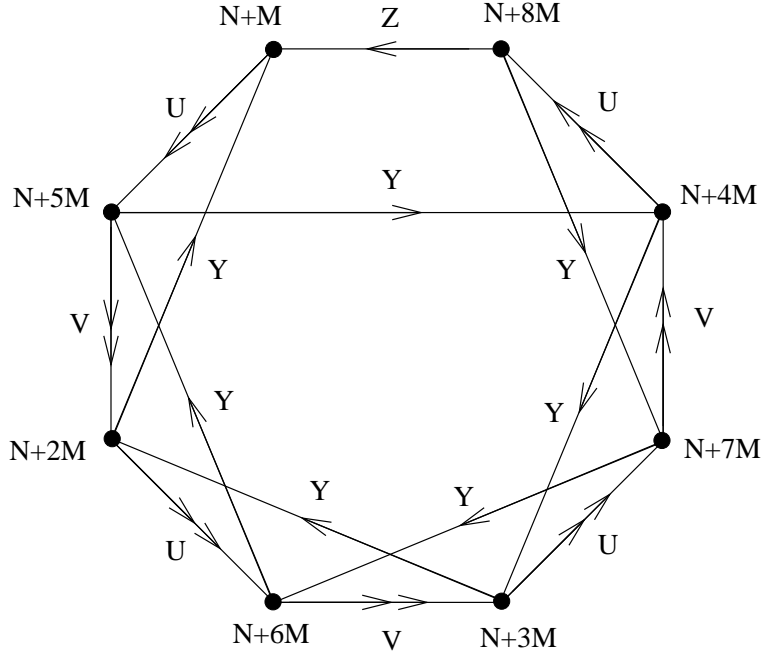}
\end{center}
\vfil
\caption{The quiver for $Y^{4,3}$ reproduced from Figure 4 of \cite{Benvenuti}.
This quiver is identical to Figure 1c.
}
\label{polyfig}
\end{figure}

As shown in 
\cite{Benvenuti}, the
systematics of the quiver diagram emerges most clearly for $p>2$ where,
placing the gauge groups at the vertices of a regular polygon,
we find that the outer edge of the diagram consists of $2p$
vertices connected by double arrows 
pointing in the same direction, except for one ``impurity'' where
the double arrow is replaced by a single one. The effect 
of the impurity is also to
merge two inner single arrows into one (see Figure \ref{polyfig}).
In the language of section 2, the $Y^{p,p-1}$ gauge theories consist
of $(p-1)$ $\sigma$ unit cells and one $\tau$ unit cell.

Upon addition of $M$ fractional branes, the single arrow ``impurity'' connects
the smallest gauge group $SU(N+M)$ with the 
biggest gauge group $SU(N+ 2p M)$.
In the case of $p=4$ corresponding to Fig. \ref{polyfig}, 
the action of the Seiberg duality on $SU(N+8M)$ gives $SU(N)$ 
because the group effectively
has $2N+ 8M$ flavors. Then we permute the adjacent vertices 
corresponding to $SU(N)$
and $SU(N+4M)$ to find a quiver identical to the  one we started with, except
with $N\to N-M$. Compared to the original diagram, the impurity moved two
steps clockwise around the outer edge.

 For the general $p$, there are $2p$ gauge groups.  On the 
conformal surface, the gauge groups are all $SU(N)$.  However,
we can add M D5-branes which shift the gauge groups to
\be
\prod_{i=1}^{2p} SU(N_i)
\ee
where
\be
N_{2n-1} = N + n M \; ; \; \; \;
N_{2n} = N + (p+n)M \ .
\ee
To be painfully explicit, the gauge group becomes
\begin{eqnarray}
&& SU(N+M) \times SU(N+(p+1)M) \times \nonumber \\ 
&& \times SU(N+2M) \times SU(N+(p+2)M) \times 
\cdots  \nonumber \\  
&& \cdots  \times SU(N+pM) \times SU(N+2pM) \ .
\end{eqnarray}

Clearly, this action of Seiberg duality generalizes to higher $p$.
The action on the biggest gauge group $SU(N+ 2p M)$ reduces it to $SU(N)$.
Subsequent permutation of adjacent vertices $SU(N)$ and $SU(N+ p M)$
turns the quiver into the one we started with, but with $N\to N-M$.

We now check that the gauge group with the most colors $SU(N+2pM)$ is
also the gauge group with the largest $\beta$ function.
All $\beta$ functions are proportional to $M$.
Setting $M=1$, we find
\begin{eqnarray}
\beta_1 &=& 3 + \frac{3}{2}
\left[ 2(p+1) (R_U -1) + 2 (R_Y-1) + 2p(R_Z-1) \right] \ , \\
\beta_{2n} &=& 3(n+p) + \frac{3}{2} \left[
2(n+1)(R_V-1) + (n+p+1)(R_Y-1) + \right. \nonumber \\
&& \left. (n+p-1)(R_Y-1) + 2n(R_U-1)
\right]
\ , \\
\beta_{2n-1} &=& 3n + \frac{3}{2} \left[
2(p+n)(R_U-1) + (n+1)(R_Y-1) + \right. \nonumber \\
&& \left. (n-1)(R_Y-1) + 2(p+n-1)(R_V-1) 
\right]
\ , \\
\beta_{2p} &=& 6p + \frac{3}{2} \left[
(R_Z-1) + (2p-1)(R_Y-1) + 2p(R_U-1) \right] \ ,
\end{eqnarray}
where for the  $\beta_{2n}$, $1 \leq n \leq p-1$
and for
the $\beta_{2n-1}$, $2 \leq n
\leq p$.

Using the fact that the superpotential has R-charge two, we see
that $R_U + R_V + R_Y = 2$
and $2R_U + R_Y + R_Z = 2$, from which it follows that
\begin{eqnarray*}
\beta_1 = -\beta_{2p} &=& \frac{3}{2} ( R_Y - R_Z - 2p(R_U + R_Y)) \ , \\
\beta_{2n-1} = -\beta_{2n} &=& 3(1-R_V - p R_Y) \ . 
\end{eqnarray*}
 From (\ref{RY}), (\ref{RZ}), (\ref{RU}), and (\ref{RV}), one can check that
\be
\beta_{1} < \beta_{2n-1} < 0 < \beta_{2n} < \beta_{2p} \ .
\ee
In particular, 
\be
\beta_1 = -5p + \sqrt{p^2 - (p-1)^2} < 0 \ ,
\ee
for $p\geq 1$.  Moreover,
consider the difference
\be
\beta_{2n-1} - \beta_{1} = \frac{2p^2 \left(2p - \sqrt{p^2+6p-3} \right)}
{(1-p)^2} \ .
\ee
This difference is strictly greater than zero for $p\geq 1$.
We conclude that $\beta_1$ and $\beta_{2p}$ have the largest magnitude
of the $2p$ $\beta$-functions. Therefore,
as the theory flows to the IR, the coupling will generically blow up
first for the biggest gauge group $SU(N+ 2 p M)$, necessitating
an application of
Seiberg duality. 

To make sure that we did not make a mistake, we
check that
\be
\beta_{D5} =  \sum s^i \beta_i = 
 \sum_{n=1}^p n \beta_{2n-1} +  \sum_{n=1}^p (p+n) \beta_{2n} \ ,
\ee
where the $s^i$ is the D5-brane vector.
Lo and behold,
\be
\sum s^i \beta_i = p^2 \left(p + \sqrt{4p^2 - 3(p-1)^2} \right) M \ ,
\ee
in agreement with (\ref{betageom}).

\subsection{Cascades for $Y^{p,1}$}

\begin{figure}
\begin{center}
\includegraphics[width=4in]{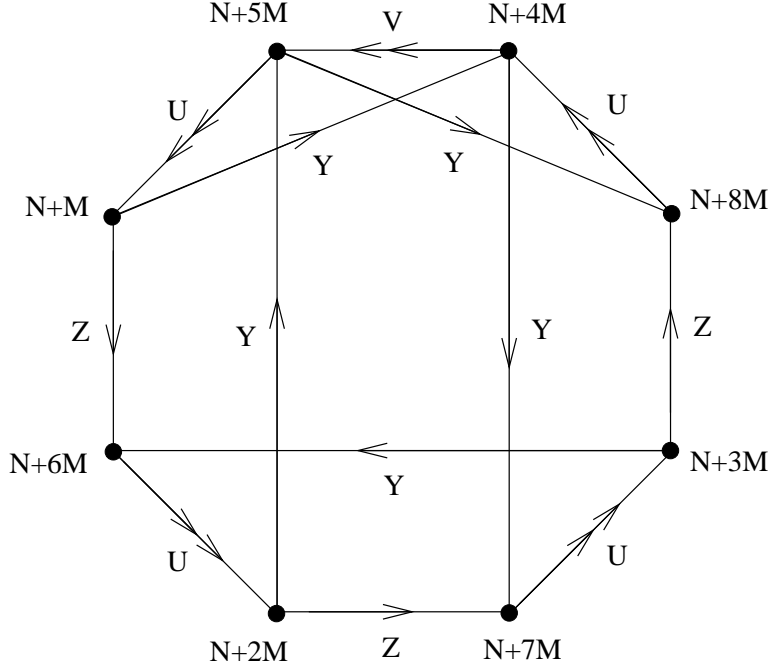}
\end{center}
\vfil
\caption{The quiver theory for $Y^{4,1}$, involving three $\tau$ unit cells and one $\sigma$
unit cell.
}
\label{Y41oct}
\end{figure}

The $Y^{2,1}$ theory is not only the simplest example of $Y^{p,p-1}$ but also of $Y^{p,1}$.
The $Y^{p,1}$ quivers, in the language of Section \ref{LeighStrassler}, contain
$(p-1)$ $\tau$ unit cells and one $\sigma$ unit cell.  The quiver for $Y^{4,1}$ is shown
as Figure \ref{Y41oct}.

The gauge groups for the $Y^{p,1}$ spaces are
\be
\prod_{i=1}^{2p} SU(N_i)
\ee
where
\be
N_{2n-1} = N + (p + n)M \; ; \; \; \; N_{2n} = N + nM \ ,
\ee
where the $\sigma$ unit cell contains both the first and second 
and also the last and second to last
gauge groups.

\begin{figure}
\begin{center}
\includegraphics[width=6in]{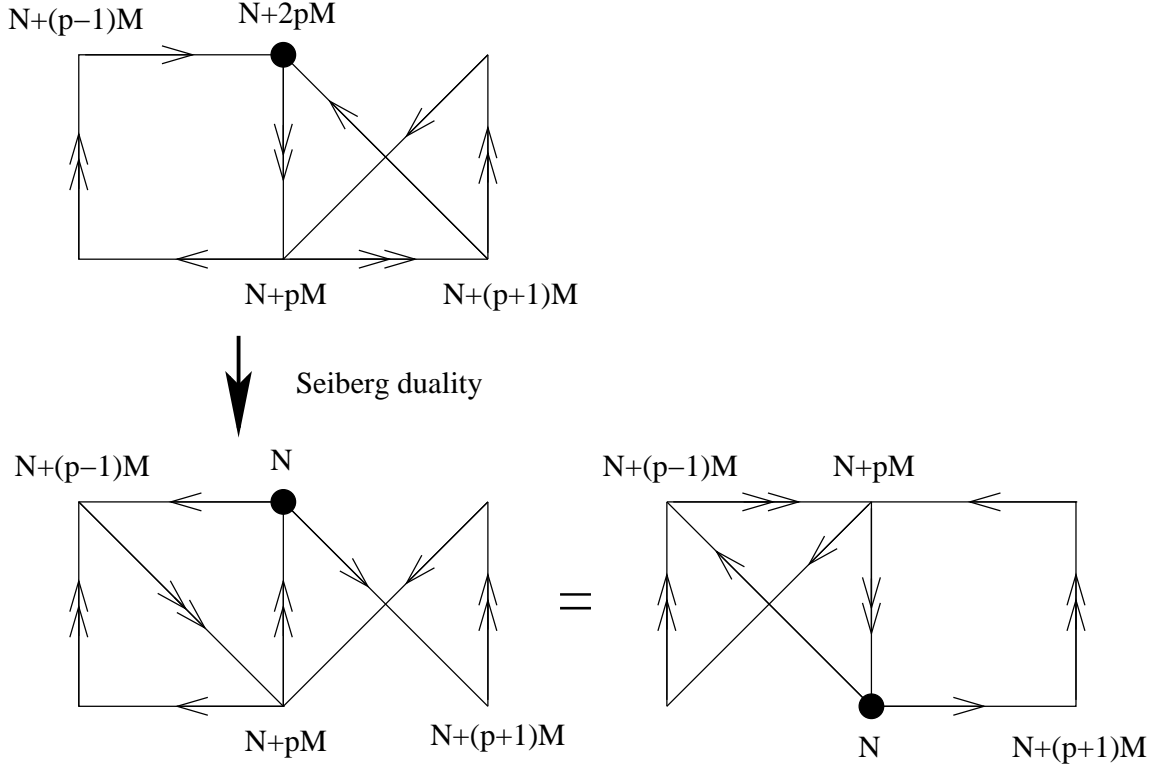}
\end{center}
\vfil
\caption{Seiberg duality for the $Y^{p,1}$ quiver: 
$(\cdots \tau \tilde \tau \sigma \tilde \tau \cdots )\to 
(\cdots \tau \tilde \sigma \tau \tilde \tau \cdots )$.
}
\label{seiberg}
\end{figure}

The gauge groups with the largest and smallest numbers of colors are associated
with the impurity, i.e.~the $\sigma$ unit cell.  The gauge group with the largest number of
colors $SU(N+2pM)$ has $2N+2pM$ flavors.  Thus, after a Seiberg duality, the
gauge group will change to $SU(N)$.  Switching this $SU(N)$ gauge group with
its neighbor $SU(N+pM)$ we find the same quiver but with the $\sigma$ 
impurity shifted one cell to the left and $N\to N-M$ (see Fig. \ref{seiberg}).

We now check whether Seiberg duality will generically happen at the
gauge group with the largest number of colors.
The $\beta$-functions for the $2p$ gauge groups are
\begin{eqnarray*}
\beta_1 = -\beta_{2p} &=& 3 (p-1 + (1-p) R_U + R_Y) M \ ,\\
\beta_{2n+1} = -\beta_{2n} &=& 3 \left( p + \frac{1}{2} R_Y - \frac{1}{2} R_Z \right)M \ ,
\end{eqnarray*}
where $1 \leq n < p$.
 From (\ref{RY}), (\ref{RZ}), (\ref{RU}), and (\ref{RV}), one can check that
\be
\beta_{2n} < \beta_{2p} < 0 < \beta_1 < \beta_{2n+1} \ .
\ee

Indeed, the gauge group with the largest number of colors has the largest
$\beta$ function.  However, an important difference between the $Y^{p,p-1}$
and the $Y^{p,1}$ gauge theories is that in the present case, there
are $p-2$ other gauge groups which share the same large $\beta$-function.
It may happen that Seiberg duality occurs first at the node with the largest
number of colors, but the situation is less generic than before.

 Finally, we check the sum
\be
\sum_{i=1}^{2p} s^i \beta_i  = \sum_{n=1}^p  (p+n) \beta_{2n-1} + \sum_{n=1}^p n \beta_{2n} 
= p^2 \left( p + \sqrt{4p^2 - 3} \right) M
\ .
\ee
This result agrees with our expectations from (\ref{betageom}).

\subsection{The Baryonic Branch}

 Both
for $Y^{p,1}$ and $Y^{p,p-1}$, if initially $N$ is a multiple of $M$ then far in the IR $N$ is reduced to zero,
so that we find the gauge group
$ SU(M)\times SU(2M) \times \ldots \times SU(2p M)$. Note that for the $SU(2pM)$ factor
there are effectively $2p M$ flavors. Hence we can form baryon operators.
In this sense the cascade obtained  
is rather analogous to
the cascade found with $T^{1,1}$ (the latter case formally 
corresponds to $p=1$ and $q=0$).
It is therefore possible that all these theories have a baryonic branch
where the $U(1)_B$ and the $U(1)_F$  
continuous symmetries are spontaneously broken. 
This idea needs further investigation because the dynamics
of the $ SU(M)\times SU(2M)\times \ldots \times SU(2p M)$ 
gauge theory is necessarily more complex
than for the $SU(M)\times SU(2M)$ case found for the deformed conifold.

\section*{Acknowledgments}
We would like to single Carlos Nu\~nez out for special thanks
for his suggestions and for his participation in the early stages
of this collaboration.
We are grateful to Sergio Benvenuti, 
Jarah Evslin, Seba Franco, Amihay Hanany, 
James Sparks,
and Brian Wecht for useful discussions.
Q.~J.~E. would also like to thank Hong Liu.
C.~P.~H. would like to thank the Harvard Physics Department, where this
project began, for hospitality.
The research of I.~R.~K. is
supported in part by the National Science Foundation Grant No.~PHY-0243680.
The research of C.~P.~H. is supported in part by the NSF under Grant No.~PHY99-07949.
The research of Q.~J.~E. is supported in part by funds provided by the U.S.D.O.E. 
under co-operative research agreement
DF-FC02-94ER40818.
Any opinions, findings, and conclusions or recommendations expressed in
this material are those of the authors and do not necessarily reflect
the views of the National Science Foundation.


\begin{thebibliography}{99}


\bibitem{jthroat}
J.~M.~Maldacena,
``The large N limit of superconformal field theories and supergravity,''
Adv.\ Theor.\ Math.\ Phys.\  {\bf 2}, 231 (1998)
[Int.\ J.\ Theor.\ Phys.\  {\bf 38}, 1113 (1999)]
[arXiv:hep-th/9711200].

\bibitem{US}
S.~S.~Gubser, I.~R.~Klebanov and A.~M.~Polyakov,
``Gauge theory correlators from non-critical string theory,''
Phys.\ Lett.\ B {\bf 428}, 105 (1998)
[arXiv:hep-th/9802109].

\bibitem{EW}
E.~Witten,
``Anti-de Sitter space and holography,''
Adv.\ Theor.\ Math.\ Phys.\  {\bf 2}, 253 (1998)
[arXiv:hep-th/9802150].

\bibitem{ks}
S.~Kachru and E.~Silverstein,
``4d conformal theories and strings on orbifolds,''
Phys.\ Rev.\ Lett.\  {\bf 80}, 4855 (1998)
[arXiv:hep-th/9802183]; \\
A.~E.~Lawrence, N.~Nekrasov and C.~Vafa,
``On conformal field theories in four dimensions,''
Nucl.\ Phys.\ B {\bf 533}, 199 (1998)
[arXiv:hep-th/9803015].

\bibitem{Kehag}
A.~Kehagias,
``New type IIB vacua and their F-theory interpretation,''
Phys.\ Lett.\ B {\bf 435}, 337 (1998)
[arXiv:hep-th/9805131].

\bibitem{KW}
I.~R.~Klebanov and E.~Witten,
``Superconformal field theory on threebranes at a Calabi-Yau  singularity,''
Nucl.\ Phys.\ B {\bf 536}, 199 (1998)
[arXiv:hep-th/9807080].

\bibitem{MP}
D.~R.~Morrison and M.~R.~Plesser,
``Non-spherical horizons. I,''
Adv.\ Theor.\ Math.\ Phys.\  {\bf 3}, 1 (1999)
[arXiv:hep-th/9810201].

\bibitem{Acharya}
B.~S.~Acharya, J.~M.~Figueroa-O'Farrill, C.~M.~Hull and B.~Spence,
``Branes at conical singularities and holography,''
Adv.\ Theor.\ Math.\ Phys.\  {\bf 2}, 1249 (1999)
[arXiv:hep-th/9808014].

\bibitem{Gauntlett2}
J.~P.~Gauntlett, D.~Martelli, J.~Sparks and D.~Waldram,
``Supersymmetric AdS(5) solutions of M-theory,''
Class.\ Quant.\ Grav.\  {\bf 21}, 4335 (2004)
[arXiv:hep-th/0402153].

\bibitem{Gauntlett}
J.~P.~Gauntlett, D.~Martelli, J.~Sparks and D.~Waldram,
``Sasaki-Einstein metrics on S(2) x S(3),''
arXiv:hep-th/0403002.

\bibitem{Martelli}
D.~Martelli and J.~Sparks,
``Toric geometry, Sasaki-Einstein manifolds and a new infinite class of
AdS/CFT duals,''
arXiv:hep-th/0411238.

\bibitem{Benvenuti}
S.~Benvenuti, S.~Franco, A.~Hanany, D.~Martelli and J.~Sparks,
``An infinite family of superconformal quiver gauge theories with
Sasaki-Einstein duals,''
arXiv:hep-th/0411264.

\bibitem{Feng}
B.~Feng, A.~Hanany and Y.~H.~He,
``D-brane gauge theories from toric singularities and toric duality,''
Nucl.\ Phys.\ B {\bf 595}, 165 (2001)
[arXiv:hep-th/0003085].


\bibitem{Bertolini}
M.~Bertolini, F.~Bigazzi and A.~L.~Cotrone,
``New checks and subtleties for AdS/CFT and a-maximization,''
arXiv:hep-th/0411249.

\bibitem{chaotic}
S.~Franco, Y.~H.~He, C.~Herzog and J.~Walcher,
``Chaotic duality in string theory,''
Phys.\ Rev.\ D {\bf 70}, 046006 (2004)
[arXiv:hep-th/0402120]; \\
``Chaotic Cascades for D-branes on Singularities,''
proceedings of
the Cargese Summer School 2004,
[arXiv:hep-th/0412207].

\bibitem{Hanwalls}
S.~Franco, A.~Hanany, Y.~H.~He and P.~Kazakopoulos,
``Duality walls, duality trees and fractional branes,''
arXiv:hep-th/0306092.

\bibitem{KT}
I.~R.~Klebanov and A.~A.~Tseytlin,
``Gravity duals of supersymmetric SU(N) x SU(N+M) gauge theories,''
Nucl.\ Phys.\ B {\bf 578}, 123 (2000)
[arXiv:hep-th/0002159].

\bibitem{KS}
I.~R.~Klebanov and M.~J.~Strassler,
``Supergravity and a confining gauge theory: Duality cascades and
chiSB-resolution of naked singularities,''
JHEP {\bf 0008}, 052 (2000)
[arXiv:hep-th/0007191].

\bibitem{cascades}
F.~Cachazo, B.~Fiol, K.~A.~Intriligator, S.~Katz and C.~Vafa,
``A geometric unification of dualities,''
Nucl.\ Phys.\ B {\bf 628}, 3 (2002)
[arXiv:hep-th/0110028]; \\
%
B.~Fiol,
``Duality cascades and duality walls,''
JHEP {\bf 0207}, 058 (2002)
[arXiv:hep-th/0205155]; \\
%
A.~Hanany and J.~Walcher,
``On duality walls in string theory,''
JHEP {\bf 0306}, 055 (2003)
[arXiv:hep-th/0301231]; \\
%
R.~Ahl Laamara, M.~Ait Ben Haddou, A.~Belhaj, L.~B.~Drissi and E.~H.~Saidi,
``RG cascades in hyperbolic quiver gauge theories,''
Nucl.\ Phys.\ B {\bf 702}, 163 (2004)
[arXiv:hep-th/0405222].

\bibitem{IW}
K.~Intriligator and B.~Wecht,
``The exact superconformal R-symmetry maximizes a,''
Nucl.\ Phys.\ B {\bf 667}, 183 (2003)
[arXiv:hep-th/0304128].

\bibitem{Leigh}
R.~G.~Leigh and M.~J.~Strassler,
``Exactly marginal operators and duality in four-dimensional N=1
supersymmetric gauge theory,''
Nucl.\ Phys.\ B {\bf 447}, 95 (1995)
[arXiv:hep-th/9503121].

\bibitem{GK}
S.~S.~Gubser and I.~R.~Klebanov,
``Baryons and domain walls in an N = 1 superconformal gauge theory,''
Phys.\ Rev.\ D {\bf 58}, 125025 (1998)
[arXiv:hep-th/9808075]; \\
D.~Berenstein, C.~P.~Herzog and I.~R.~Klebanov,
``Baryon spectra and AdS/CFT correspondence,''
JHEP {\bf 0206}, 047 (2002)
[arXiv:hep-th/0202150].

\bibitem{review}
C.~P.~Herzog, I.~R.~Klebanov and P.~Ouyang,
``D-branes on the conifold and N = 1 gauge / gravity dualities,''
arXiv:hep-th/0205100; \\
C.~P.~Herzog, I.~R.~Klebanov and P.~Ouyang,
``Remarks on the warped deformed conifold,''
arXiv:hep-th/0108101.

\bibitem{Bert}
M.~Bertolini, P.~Di Vecchia, M.~Frau, A.~Lerda, R.~Marotta and I.~Pesando,
``Fractional D-branes and their gauge duals,''
JHEP {\bf 0102}, 014 (2001)
[arXiv:hep-th/0011077].

\bibitem{KN}
I.~R.~Klebanov and N.~A.~Nekrasov,
``Gravity duals of fractional branes and logarithmic RG flow,''
Nucl.\ Phys.\ B {\bf 574}, 263 (2000)
[arXiv:hep-th/9911096].

\bibitem{Joe}
J.~Polchinski,
``N = 2 gauge-gravity duals,''
Int.\ J.\ Mod.\ Phys.\ A {\bf 16}, 707 (2001)
[arXiv:hep-th/0011193].

\bibitem{Dasgupta}
K.~Dasgupta, G.~Rajesh and S.~Sethi,
``M theory, orientifolds and G-flux,''
JHEP {\bf 9908}, 023 (1999)
[arXiv:hep-th/9908088]; \\
M.~Grana and J.~Polchinski,
``Supersymmetric three-form flux perturbations on AdS(5),''
Phys.\ Rev.\ D {\bf 63}, 026001 (2001)
[arXiv:hep-th/0009211]; \\
S.~S.~Gubser,
``Supersymmetry and F-theory realization of the deformed conifold with
three-form flux,''
arXiv:hep-th/0010010.

\bibitem{Halmagyi}
N.~Halmagyi, K.~Pilch, C.~Romelsberger and N.~P.~Warner,
``The complex geometry of holographic flows of quiver gauge theories,''
arXiv:hep-th/0406147.

\bibitem{Seibergduality}
N.~Seiberg, 
``Electric - magnetic duality in supersymmetric nonAbelian gauge theories,''
Nucl.\ Phys.\ B {\bf 435}, 129 (1995)
[arXiv:hep-th/9411149].

\bibitem{Aharony}
O.~Aharony,
``A note on the holographic interpretation of string theory backgrounds  with
varying flux,''
JHEP {\bf 0103}, 012 (2001)
[arXiv:hep-th/0101013].

\bibitem{GHK}
S.~S.~Gubser, C.~P.~Herzog and I.~R.~Klebanov,
``Symmetry breaking and axionic strings in the warped deformed conifold,''
JHEP {\bf 0409}, 036 (2004)
[arXiv:hep-th/0405282]; \\
``Variations on the warped deformed conifold,''
arXiv:hep-th/0409186.


\bibitem{Butti}
A.~Butti, M.~Grana, R.~Minasian, M.~Petrini and A.~Zaffaroni,
``The Baryonic Branch of Klebanov-Strassler Solution: a Supersymmetric Family
of SU(3) Structure Backgrounds,''
arXiv:hep-th/0412187.


\end{thebibliography}
\end{document}